\documentclass[11pt]{article}
\usepackage{amsmath,amssymb,amsthm,bbm,epsfig}
\usepackage[sc]{mathpazo}

\usepackage{array}
\newcommand{\PreserveBackslash}[1]{\let\temp=\\#1\let\\=\temp}
\newcolumntype{C}[1]{>{\PreserveBackslash\centering}p{#1}}
\newcolumntype{R}[1]{>{\PreserveBackslash\raggedleft}p{#1}}
\newcolumntype{L}[1]{>{\PreserveBackslash\raggedright}p{#1}}

\begin{document}

\title{\textbf{$\mu$-$\tau$ reflection symmetry and radiative corrections}}
\author{\\[.5mm]Ye-Ling Zhou \footnote{E-mail: {\tt zhouyeling@ihep.ac.cn}}\\ \\
{\it\small Institute of High Energy Physics, Chinese Academy of Sciences, Beijing 100049, China}
}
\date{}

\maketitle

\begin{abstract}
The $\mu$-$\tau$ reflection symmetry is compatible with current neutrino oscillation data and easily realized under family symmetries. We prove that this symmetry preserves $\theta_{23}=45^\circ$, $\delta=\pm90^\circ$, $\rho,\sigma=0,90^\circ$, and can be embedded into the seesaw mechanism. 
The $\mu$-$\tau$ reflection symmetry preserved at a high energy scale $\Lambda_\text{FS}$ will be broken by radiative corrections and result in deviations of $\theta_{23}$ from $45^\circ$ and $\delta$ from $\pm90^\circ$ at the electroweak scale. We develop an analytical method to derive the corrections to all the mixing parameters. We perform a numerical analysis in the MSSM for $\delta=-90^\circ$ at $\Lambda_\text{FS}$, and observe that $\theta_{23}>45^\circ$ in the normal mass ordering, $\theta_{23}<45^\circ$ in the inverted mass ordering, and the sizable correction to $\delta$ prefers a negative sign. These deviations have definite directions and can be tested in the future neutrino oscillation experiments.

\end{abstract}

\begin{flushleft}
\hspace{0.8cm} PACS number(s): 14.60.Pq, 13.15.+g, 25.30.Pt
\end{flushleft}

\def\thefootnote{\arabic{footnote}}
\setcounter{footnote}{0}

\newpage

\section{Introduction}

Although neutrino oscillation experiments have greatly developed our knowledge of neutrino masses and lepton flavor mixing \cite{PDG}, there are still some mysteries: the neutrino mass ordering (normal $m_1<m_2<m_3$ or inverted $m_3<m_1<m_2$), the octant of the atmospheric mixing angle $\theta_{23}$ ($\theta_{23}<45^\circ$ or $\theta_{23}>45^\circ$) and the value of the Dirac CP-violating phase $\delta$. The undergoing and upcoming neutrino oscillation experiments aim to solve these problems.

Physicists have made much effort for understanding the lepton flavor mixing.
A $\mu$-$\tau$ exchange symmetry under the transformation $\nu_{\mu\text{L}}\leftrightarrow\nu_{\tau\text{L}}$ is often assumed \cite{mutau0}. Under this symmetry,  the Majorana neutrino mass matrix takes the form 
\begin{eqnarray}
M^{}_{\nu}=
\left(
\begin{array}{ccc}
 a\;\, & \;b\; & \;\,b \\
 b\;\, & \;c\; & \;\,d \\
 b\;\, & \;d\; & \;\,c
\end{array}
\right) 
\label{mutau0}
\end{eqnarray}
in the flavor basis where the charged lepton mass matrix is diagonal. It simultaneously results in $\theta_{23}=45^\circ$ and $\theta_{13}=0$. More symmetries imposed on the mass texture in Eq.~\eqref{mutau0} can lead to the bimaximal mixing \cite{BM}, tri-bimaximal mixing \cite{TBM}, et al.

Combining the $\mu$-$\tau$ exchange symmetry with CP symmetry, we achieve the following texture:
\begin{eqnarray}
M^{}_{\nu}=
\left(
\begin{array}{ccc}
 a\;\, & \;b\;\; & b^*\! \\
 b\;\, & \;c\;\; & d\;\; \\
 b^*\!\! & \;d\;\; & c^*\!
\end{array}
\right) \,.
\label{mutau}
\end{eqnarray}
It was first suggested and realized in the family symmetry $A_4$ by Babu, Ma, and Valle \cite{mutau1}. This texture is invariant under a combination of the $\mu$-$\tau$ exchange and CP conjugate transformations \cite{Grimus:2003yn}: 
\begin{eqnarray}
\nu^{}_{e\text{L}}\to\nu_{e\text{L}}^\text{c}\,, \quad\nu^{}_{\mu\text{L}}\to\nu_{\tau\text{L}}^\text{c}\,, \quad\nu^{}_{\tau\text{L}}\to\nu_{\mu\text{L}}^\text{c}\,,
\label{mutau_reflection}
\end{eqnarray}
which is regarded as a typical kind of generalized CP transformations \cite{Feruglio:2012cw,Holthausen:2012dk}. In Ref.~\cite{Harrison:2002et}, Harrison and Scott gave it the name ``$\mu$-$\tau$ reflection''. Although only Dirac neutrinos were assumed in their original paper, the concept of  $\mu$-$\tau$ reflection has been inherited and used in the literature, e.g., see \cite{Feruglio:2012cw, Farzan:2006vj}. In Ref.~\cite{Zhou:2012zj}, it is also called the  generalized $\mu$-$\tau$ transformation.
The observation of a sizable reactor angle $\theta_{13}\simeq8.8^\circ$ \cite{DYB} and the hint for the maximal CP violation $\delta\sim-90^\circ$ \cite{T2K} indicate that the $\mu$-$\tau$ reflection symmetry may be an approximate symmetry in the neutrino sector \cite{Xing:2014zka}. Later we will prove that $\theta_{23}=45^\circ$, $\delta=\pm90^\circ$, and $\rho,\sigma=0,90^\circ$ must be required by the $\mu$-$\tau$ reflection symmetry, and $\theta_{12}$ and $\theta_{13}$ are left arbitrary. One can further constrain $\theta_{12}$ and $\theta_{13}$ by requiring more relations, e.g., the connection with TM$_1$ and TM$_2$ \cite{TM}. More discussions on the $\mu$-$\tau$ reflection symmetry in the general case can be found in Ref.~\cite{more}.

The $\mu$-$\tau$ reflection symmetry can be realized under family symmetries. Typically in the framework of generalized CP symmetries \cite{Feruglio:2012cw}, it is easily realized by requiring a combination of the family symmetry and CP symmetry $G_\text{f}\rtimes CP$ breaking to remnant symmetries $Z_n$ and $Z_2\times CP$ in the charged lepton and neutrino sectors, respectively. In addition, $\theta_{12}$ and $\theta_{13}$ are constrained and dependent upon a single parameter. There are a lot of model-independent analyses of how to derive this symmetry in $A_4$ \cite{Ding:2013bpa}, $S_4$ \cite{Holthausen:2012dk,Ding:2013hpa}, $\Delta(48)$ \cite{D48}, and $\Delta(96)$ \cite{D96}. For explicit models constructed in generalized CP, please see \cite{Ding:2013bpa,Ding:2013hpa, Feruglio:2013hia, models}. The Friedberg-Lee symmetry can also lead to this mass texture and constrain the mixing angles $\theta_{12}$ and $\theta_{13}$ \cite{FL}. 

The renormalization group (RG) running effect will contribute to the neutrino mass matrix and modify mass eigenvalues and mixing parameters \cite{RGE1,RGE2,RGE3,Antusch:2001vn}.
Even if the $\mu$-$\tau$ reflection symmetry is explicitly preserved at a high energy scale, it must be broken due to the RG equations running down to a low energy scale. And the mixing angle $\theta_{23}$ and the Dirac phase $\delta$ deviate from $45^\circ$ and $\pm90^\circ$, respectively. Recently, the RG running effect of a $\mu$-$\tau$ symmetry at the PMNS matrix level has been shown schematically \cite{Luo:2014upa}. In their paper, the assumption of the PMNS matrix elements $|U_{\mu i}|=|U_{\tau i}|$ (for $i=1,2,3$) has been made, which results in $\theta_{23}=45^\circ$, $\delta=\pm90^\circ$ but leaves $\rho$, $\sigma$ arbitrary. 
In the present paper, we will give a general discussion of $\mu$-$\tau$ reflection symmetry and an analytical description of its RG running effects. 

The rest of our paper is organized as follows. Section 2 is devoted to the basic feature of flavor mixing in the $\mu$-$\tau$ reflection symmetry and an extended discussion of how to embed it to the seesaw mechanism. In section 3, we systematically analyze the RG running effects in both analytical and numerical approaches. In general, these effects can be divided into two parts: $\mu$-$\tau$ symmetric and anti-symmetric. We summarize our results in section 4. 

\section{$\mu$-$\tau$ reflection symmetry}

\subsection{Flavor mixing}

Given any neutrino mass matrix $M_{\nu}$ in the form of Eq.~\eqref{mutau} that preserves the $\mu$-$\tau$ reflection symmetry, we do the following transformation 
\begin{eqnarray}
U^\dag_{23} M_{\nu}U^*_{23} = 
\left(
\begin{array}{ccc}
 a & \sqrt{2}\text{Im}(b) & \sqrt{2}\text{Re}(b) \\
 \sqrt{2}\text{Im}(b) & d-\text{Re}(c) & \text{Im}(c) \\
 \sqrt{2}\text{Re}(b) & \text{Im}(c) & d+\text{Re}(c)
\end{array}
\right)
\label{real_mass_matrix}
\end{eqnarray}
with
\begin{eqnarray}
U^{}_{23} = 
\left(
\begin{array}{ccc}
 1 & 0 & 0 \\
 0 & \frac{i}{\sqrt{2}} & \frac{1}{\sqrt{2}} \\
 0 & \frac{-i}{\sqrt{2}} & \frac{1}{\sqrt{2}}
\end{array}
\right)
\end{eqnarray}
and ``Re'' and ``Im'' denoting the real and imaginary parts, respectively. We see that the RHS of Eq.~\eqref{real_mass_matrix} is a real matrix. It can be diagonalized by a real orthogonal matrix $O$ with
\begin{eqnarray}
O=
\left(
\begin{array}{ccc}
 \eta_1 & 0 & 0 \\
 0 & \eta_2 & 0 \\
 0 & 0 & \eta_3
\end{array}
\right)\left(
\begin{array}{ccc}
 1 & 0 & 0 \\
 0 & c_{1} & s_{1} \\
 0 & -s_{1} & c_{1}
\end{array}
\right)\left(
\begin{array}{ccc}
 c_{2} & 0 & s_{2} \\
 0 & 1 & 0 \\
 -s_{2} & 0 & c_{2}
\end{array}
\right)\left(
\begin{array}{ccc}
 c_{3} & s_{3} & 0 \\
 -s_{3} & c_{3} & 0 \\
 0 & 0 & 1
\end{array}
\right)\,,
\end{eqnarray}
in which $c_{i}=\cos\theta_{i}$ and $s_{i}=\sin\theta_{i}$. Here $\eta_{1,2,3}=\pm1$ are used to guarantee $0\leqslant\theta_{i}\leqslant90^\circ$. The diagonalized mass matrix can be presented by 
\begin{eqnarray}
O^T_{} U^\dag_{23} M_{\nu}U^*_{23} O=\widehat{M}_\nu\equiv\eta'\text{diag}\{\eta_\rho m_{1}, ~\eta_\sigma m_{2},~m_{3} \}\,,
\end{eqnarray} 
where $m_i$ are the absolute neutrino masses in the neutrino mass eigenstates, and $\eta',\eta_{\rho,\sigma}=\pm1$ are used to guarantee positive masses $m_i\geqslant0$. 

Based on the above discussion, we can derive the PMNS matrix which is compatible with $\mu$-$\tau$ reflection symmetry 
\begin{eqnarray}
&&U(\Lambda_\text{FS})=U_{23}O\, \sqrt{\eta'} \text{diag}\{\sqrt{\eta_\rho}, ~\sqrt{\eta_\sigma} ,~1\}\nonumber\\
&&= \eta_3\sqrt{\eta'}
\left(
\begin{array}{ccc}
 \frac{\eta_1}{\eta_3} & 0 & 0 \\
 0 & e^{i\theta_{1}} & 0 \\
 0 & 0 & e^{-i\theta_{1}}
\end{array}
\right)\left(
\begin{array}{ccc}
 1 & 0 & 0 \\
 0 & \frac{1}{\sqrt{2}} & \frac{1}{\sqrt{2}} \\
 0 & \frac{-1}{\sqrt{2}} & \frac{1}{\sqrt{2}}
\end{array}
\right)\left(
\begin{array}{ccc}
 c_{2} & 0 & s_{2} \\
 0 & -i\eta & 0 \\
 -s_{2} & 0 & c_{2}
\end{array}
\right)\left(
\begin{array}{ccc}
 c_{3} & s_{3} & 0 \\
 -s_{3} & c_{3} & 0 \\
 0 & 0 & 1
\end{array}
\right)\left(
\begin{array}{ccc}
 \sqrt{\eta_\rho} & 0 & 0 \\
 0 & \sqrt{\eta_\sigma} & 0 \\
 0 & 0 & 1
\end{array}
\right)\,,
\label{PMNSofmutau}
\end{eqnarray}
where $\eta_\delta=-\eta_2/\eta_3=\pm1$. We use the convention of the PMNS matrix as follows:
\begin{eqnarray}
U=\left(
\begin{array}{ccc}
 1 & 0 & 0 \\
 0 & c_{23} & s_{23} \\
 0 & -s_{23} & c_{23}
\end{array}
\right)\left(
\begin{array}{ccc}
 c_{13} & 0 & s_{13} \\
 0 & e^{-i\delta } & 0 \\
 -s_{13}  & 0 & c_{13}
\end{array}
\right)\left(
\begin{array}{ccc}
 c_{12} & s_{12} & 0 \\
 -s_{12} & c_{12} & 0 \\
 0 & 0 & 1
\end{array}
\right)\left(
\begin{array}{ccc}
 e^{i\rho} & 0 & 0 \\
 0 & e^{i\sigma} & 0 \\
 0 & 0 & 1
\end{array}
\right)\,,
\end{eqnarray}
where $c_{ij}=\cos\theta_{ij}$, $s_{ij}=\sin\theta_{ij}$, $\delta$ is the Dirac CP-violating phase, $\rho$, $\sigma$ are Majorana CP-violating phases, and $-180^\circ\leqslant\delta<180^\circ$, $0\leqslant\rho,\sigma<180^\circ$ are required. Comparing Eq.~\eqref{PMNSofmutau} with this convention and ignoring the unphysical phases, we obtain the predicted lepton mixing parameters 
\begin{eqnarray}
&&\theta_{23}=45^\circ\,,\qquad\delta=\eta_\delta 90^\circ =\pm 90^\circ\,,
\qquad \theta_{12}=\theta_{3}\,, \qquad\theta_{13}=\theta_{2}\,, \nonumber\\
&&\rho=\arg\sqrt{\eta_\rho}= \left\{\begin{array}{lll}
 0,&&\eta_\rho=+1\\
 90^\circ,&&\eta_\rho=-1
\end{array}
\right.\,,\qquad
\sigma=\arg\sqrt{\eta_\sigma}=\left\{\begin{array}{lll}
 0,&&\eta_\sigma=+1\\
 90^\circ,&&\eta_\sigma=-1
\end{array}
\right.\,,
\end{eqnarray}
exactly. We see that if the neutrino mass matrix maintains the $\mu$-$\tau$ reflection symmetry, $\theta_{23}$ and all the CP-violating phases $\delta$, $\rho$, $\sigma$ take definite values. The parameters $\eta_\delta,\eta_\rho,\eta_\sigma=\pm1$ have physical meaning and the two discrete values $\pm1$  cannot be determined by the symmetry. 
Finally, $\theta_1$ becomes an unphysical phase which can be rotated away by redefinition of the phases of charged leptons. 

We would like to emphasize the phenomenological importance of the $\mu$-$\tau$ reflection symmetry. 
The atmospheric mixing angle $\theta_{23}=45^\circ$ and the Dirac phase $\delta=-90^\circ$ are not far away from their best-fit values of current global-fit data of neutrino oscillations, and they keep unchanged when constant matter effects are taken into account for long-baseline neutrino oscillation experiments  due to the Toshev relation \cite{Toshev:1991ku,mattereffect}. The Majorana phases $\rho,\sigma$ are fixed at $0$ or $90^\circ$, which reduce the parameter space of the effective neutrino mass term $\langle m\rangle_{ee}$ in neutrinoless double-beta decay experiments.

\subsection{$\mu$-$\tau$ reflection under the seesaw mechanism}

To explain tiny neutrino masses, we take account of the type-I seesaw mechanism. This subsection is devoted to an approach which combines the $\mu$-$\tau$ reflection symmetry  with the type-I seesaw mechanism. We give the neutrino mass terms in the basis where the charged lepton mass matrix is diagonal, 
\begin{eqnarray}
-\mathcal{L}_\text{mass}=\overline{\nu_{\text{L}}} M_\text{D}N_{\text{R}}
+ \frac{1}{2} \overline{N_{\text{R}}^\text{c}} M_\text{R}N_{\text{R}}^{}
+\text{h.c.}\,,
\label{lagrangian}
\end{eqnarray}
in which $\nu^{}_\text{L}=(\nu^{}_{e\text{L}},\nu^{}_{\mu\text{L}},\nu^{}_{\tau\text{L}})^T$, $N^{}_\text{R}=(N^{}_{x\text{R}},N^{}_{y\text{R}},N^{}_{z\text{R}})^T$ are the left-handed and right-handed neutrinos, respectively. The extended $\mu$-$\tau$ reflection transformation can be defined as
\begin{eqnarray}
&&\nu^{}_{e\text{L}}\to\nu_{e\text{L}}^\text{c}\,, \quad\;\;\; \nu^{}_{\mu\text{L}}\to\nu_{\tau\text{L}}^\text{c}\,, \quad\;\; \nu^{}_{\tau\text{L}}\to\nu_{\mu\text{L}}^\text{c}\,, \nonumber\\
&&N^{}_{x\text{R}}\to N_{x\text{R}}^\text{c}\,, \quad N^{}_{y\text{R}}\to N_{z\text{R}}^\text{c}\,, \quad N^{}_{z\text{R}}\to N_{y\text{R}}^\text{c}\,.
\label{mutau_reflection_seesaw}
\end{eqnarray}
This definition is not unique in the type-I seesaw mechanism. One can assume another generalized CP transformation different from that in Eq.~\eqref{mutau_reflection_seesaw} in the right-handed neutrino sector $N_\text{R}$. The different generalized CP transformation just corresponds to a different choice of the right-handed neutrino flavor basis, if there is no other special flavor structure imposed on the right-handed neutrino sector. 

The Dirac mass matrix $M_\text{D}$ and the right-handed Majarana mass matrix $M_\text{R}$ invariant under the above transformation must take the following forms:
\begin{eqnarray}
M^{}_\text{D}\equiv
\left(
\begin{array}{ccc}
 {\bf a}\;\, & \;{\bf b}\;\; & {\bf b}^*\! \\
 {\bf b}'\;\, & \;{\bf c}\;\; & {\bf d}\;\; \\
 {\bf b}'^*\!\! & \;{\bf d}^* & {\bf c}^*\!
\end{array}
\right) \,,~\quad
M^{}_\text{R}\equiv
\left(
\begin{array}{ccc}
 A\;\, & \;B\;\; & B^*\! \\
 B\;\, & \;C\;\; & D\;\; \\
 B^*\!\! & \;D\;\; & C^*\!
\end{array}
\right) \,.
\label{mutau_seesaw}
\end{eqnarray}
Note that $M_\text{D}$ is not necessarily a symmetric matrix, where only ${\bf a}$ is real and ${\bf b}$, ${\bf b}'$, ${\bf c}$, ${\bf d}$ are complex. $M_\text{R}$ is a symmetric matrix in the same form as in Eq.~\eqref{mutau}, in which both $A$, $D$ are real and $B$, $C$ are complex. The mass textures in Eq.~\eqref{mutau_seesaw} are the most general form invariant under the extended $\nu$-$\tau$ reflection transformation. In the case that $M_\text{R}\gg M_\text{D}$, we integrate out right-handed neutrinos and obtain the tiny masses for the left-handed neutrinos through the seesaw mechanism $M_\nu=-M_\text{D} M_\text{R} M^T_\text{D}$.

In the following, we will prove that $M_\nu$ satisifies the $\mu$-$\tau$ reflection symmetry. Applying a similar transformation as shown in Eq.~\eqref{real_mass_matrix}, we derive
\begin{eqnarray}
&&U^\dag_{23} M_\text{D}U^*_{23} = 
\left(
\begin{array}{ccc}
 {\bf a} & \sqrt{2}\text{Im}({\bf b}) & \sqrt{2}\text{Re}({\bf b}) \\
 \sqrt{2}\text{Im}({\bf b}') & \text{Re}({\bf d})-\text{Re}({\bf c}) & \text{Im}({\bf c})+\text{Im}({\bf d}) \\
 \sqrt{2}\text{Re}({\bf b}') & \text{Im}({\bf c})-\text{Im}({\bf d}) & \text{Re}({\bf d})+\text{Re}({\bf c})
\end{array}
\right)\,,\nonumber\\
&&U^\dag_{23} M_\text{R}U^*_{23} = 
\left(
\begin{array}{ccc}
 A & \sqrt{2}\text{Im}(B) & \sqrt{2}\text{Re}(B) \\
 \sqrt{2}\text{Im}(B) & D-\text{Re}(C) & \text{Im}(C) \\
 \sqrt{2}\text{Re}(B) & \text{Im}(C) & D+\text{Re}(C)
\end{array}
\right)\,.
\label{real_mass_matrix_seesaw}
\end{eqnarray}
Since both $U^\dag_{23} M_\text{D}U^*_{23}$ and $U^\dag_{23} M_\text{R}U^*_{23}$ are real, they can be diagonalized by real orthogonal matrices: 
\begin{eqnarray}
&&O^T_\text{D} U^\dag_{23} M_\text{D}U^*_{23} O'_\text{D}=\widehat{M}_\text{D}\equiv\text{diag}\{{\bf k}_1 {\bf m}_{1}, ~{\bf k}_2 {\bf m}_{2},~{\bf k}_3{\bf m}_{3} \} \,,\nonumber\\
&&O^T_\text{R} U^\dag_{23} M_\text{R}U^*_{23} O_\text{R}=\widehat{M}_\text{R}\equiv\text{diag}\{K_1 M_{1}, ~K_2 M_{2},~K_3 M_{3} \}\,,
\end{eqnarray} 
where $O_\text{D}$, $O'_\text{D}$, $O_\text{R}$ are real orthogonal matrices, $O'_\text{D}$ is not necessarily equal to $O_\text{D}$ since $U^\dag_{23} M_\text{D}U^*_{23} $ may not be symmetric, ${\bf m}_i$, $M_i$ are the absolute neutrino masses in the neutrino mass eigenstates, and ${\bf k}_i, K_i=\pm1$ are used to guarantee the mass eigenvalues ${\bf m}_i$, $M_i$ to be positive, respectively. Finally, we arrive at $U^\dag_{23}M_\nu U_{23}^*=M'$ with
\begin{eqnarray}
M'=-\big[O_\text{D} \widehat{M}_\text{D} O'_\text{D}O_\text{R}^{-1}\big] \widehat{M}^{-1}_\text{R}\big[O_\text{D} \widehat{M}_\text{D} O'_\text{D}O_\text{R}^{-1}\big]^T \,.
\end{eqnarray} 
Since $M'$ is a real symmetric matrix, we can follow the procedure in the above and  affirm that the $\mu$-$\tau$ reflection symmetry is preserved in $M_\nu$. 

\section{RG running effects of $\mu$-$\tau$ reflection symmetry}

\subsection{General formulism}

We assume that the $\mu$-$\tau$ interchange symmetry in the neutrino sector is explicitly preserved as a remnant symmetry after a certain flavor symmetry breaks at a sufficiently high energy scale $\Lambda\sim \Lambda_\text{FS}$. The neutrino mass matrix takes the form
\begin{eqnarray}
M^{}_\nu(\Lambda_\text{FS})=M_{\text{sym},0}\equiv\left(
\begin{array}{ccc}
 a^{}_0 & b^{}_0 & b^*_0 \\
 b^{}_0 & c^{}_0 & d^{}_0 \\
 b^*_0 & d^{}_0 & c^*_0
\end{array}
\right)
\label{mutau@flasy}
\end{eqnarray}
in the flavor basis, in which $a^{}_0$, $d^{}_0$ are real and $b^{}_0$, $c^{}_0$ are complex parameters. Without specified, any parameter $p_0$ or $p_{\star,0}$ in this paper stands for the running value at the scale $\Lambda_\text{FS}$.

The RG equations of neutrino masses correct the structure of the neutrino mass matrix and break the $\mu$-$\tau$ reflection symmetry when the energy scale comes down. We write out the neutrino mass matrix at the electroweak scale $\Lambda_\text{EW}$ in the integral form \cite{RGE3}
\begin{eqnarray}
M^{}_\nu(\Lambda_\text{EW})=
I_\alpha\left(
\begin{array}{ccc}
 I_e & 0 & 0 \\
 0 & I_\mu & 0 \\
 0 & 0 & I_\tau
\end{array}
\right)
M_\nu(\Lambda_\text{FS})
\left(
\begin{array}{ccc}
 I_e & 0 & 0 \\
 0 & I_\mu & 0 \\
 0 & 0 & I_\tau
\end{array}
\right)\,,
\label{RGE}
\end{eqnarray}
where
\begin{eqnarray}
I_\alpha&=&\text{exp}\left[-\frac{1}{16\pi^2} \int^{\text{ln}\Lambda_\text{FS}}_{\text{ln}\Lambda_\text{EW}} \alpha(t) \text{d}t \right]\,, \nonumber\\
I_l&=&\text{exp}\left[-\frac{C}{16\pi^2} \int^{\text{ln}\Lambda_\text{FS}}_{\text{ln}\Lambda_\text{EW}} y^2_l(t) \text{d}t \right]\,.
\end{eqnarray}
In the SM and the minimal supersymmetric model (MSSM), $C$ and $\alpha$ are given by 
\begin{eqnarray}
&&C_\text{SM}=-\frac{3}{2}\,, \qquad \alpha_\text{SM}\approx-3g^2_2+\lambda+6y^2_t\,, \nonumber\\
&&C_\text{MSSM}=1\,, \qquad \alpha_\text{MSSM}\approx-\frac{6}{5}g^2_1-6g^2_2+6y^2_t\,,
\label{RGcoefficients}
\end{eqnarray}
respectively, where $g_{1,2}$ denote the gauge couplings, $\lambda$ denotes the quartic Higgs coupling in the SM, and $y_t$, $y_l$ (for $l=e,\mu,\tau$) are Yukawa couplings of the top quark and charged leptons, respectively.

We see that in Eq.~\eqref{RGE}, $I_\alpha$ is an overall factor affecting the magnitudes of the absolute neutrino masses, and $I_l$ are flavor-dependent corrections which may modify the mass structure and flavor mixing. Due to the different signs of $C$ in Eq.~\eqref{RGcoefficients}, the flavor-dependent corrections go to opposite directions in the SM and MSSM. The Yukawa couplings $y_e$, $y_\mu$ are too small as compared with $y_\tau$, and thus $I_e$ and $I_\mu$ can be approximately set to be identities. We parameterize $I_\tau$ as $1+\epsilon$, where
\begin{eqnarray}
\epsilon&=&I_\tau-1\approx-\frac{C}{16\pi^2} \int^{\text{ln}\Lambda_\text{FS}}_{\text{ln}\Lambda_\text{EW}} y^2_\tau(t) \text{d}t \approx -\frac{C}{16\pi^2} y^2_{\tau, \text{EW}} \text{ln}\frac{\Lambda_\text{FS}}{\Lambda_\text{EW}}
\end{eqnarray}
with $y_{\tau, \text{EW}}$ being the $\tau$-lepton Yukawa coupling at the electroweak scale. 
With the help of this parametrization, we can divide $M^{}_\nu(\Lambda_\text{EW})$ into two parts: the $\mu$-$\tau$ symmetric part $M_\text{sym}$ and the $\mu$-$\tau$ anti-symmetric part $M_\text{asym}$, i.e., 
\begin{eqnarray}
&&M^{}_\nu(\Lambda_\text{EW})=M_\text{sym}+M_\text{asym}\,,\nonumber\\
&&M_\text{sym}=I_\alpha\left(
\begin{array}{ccc}
 a^{}_0 & (1+\epsilon/2)\,b^{}_0 & (1+\epsilon/2)\,b^*_0 \\
 (1+\epsilon/2)\,b^{}_0 & (1+\epsilon)\,c^{}_0 & (1+\epsilon)\,d^{}_0 \\
 (1+\epsilon/2)\,b^*_0 & (1+\epsilon)\,d^{}_0 & (1+\epsilon)\,c^*_0
\end{array}
\right)
\equiv
\left(
\begin{array}{ccc}
 a\;\, & \;b\;\; & b^*\! \\
 b\;\, & \;c\;\; & d\;\; \\
 b^*\!\! & \;d\;\; & c^*\!
\end{array}
\right) \,, \nonumber\\
&&M_\text{asym}=I_\alpha\frac{\epsilon }{2}\left(
\begin{array}{ccc}
 0 & -b^{}_0 & b^*_0 \\
 -b^{}_0 & -2c^{}_0 & 0 \\
 b^*_0 & 0 & 2c^*_0
\end{array}
\right)
=\frac{\epsilon }{2}\left(
\begin{array}{ccc}
 0 & -b & b^* \\
 -b & -2c & 0 \\
 b^* & 0 & 2c^*
\end{array}
\right)+\mathcal{O}(\epsilon^2) \,.
\end{eqnarray}
In the following, we will regard $\epsilon$ as a small parameter and establish the corrections to neutrino masses and mixing parameters through perturbation theory. Only the leading corrections in $\epsilon$ will be listed analytically. The $\mu$-$\tau$ symmetric and anti-symmetric contributions will be calculated separately.

\subsection*{$\mu$-$\tau$ symmetric corrections}
\indent 

Since $M_\text{sym}$ guarantees the $\mu$-$\tau$ reflection symmetry, the specific values $\theta_{23}=45^\circ$, $\delta=\pm90^\circ$, and $\rho, \sigma=0,90^\circ$ keep unchanged. We can use the diagonalization method given in section 2 for both $M_\text{sym}$ and $M_{\text{sym},0}$. However, the RG running effect modifies the mixing angles $\theta_{12}$, $\theta_{13}$ and the absolute neutrino masses $m_{i}$ at the energy scale $\Lambda_\text{EW}$  from their original values $\theta_{13,0}$, $\theta_{12,0}$, and $m_{i,0}$ at the scale $\Lambda_\text{FS}$. To see the connections of these parameters between two energy scales, we first apply the transformation of $U^{}_{23}$ and get two real mass matrices $U^\dag_{23} M_\text{sym}U^*_{23}$ and $U^\dag_{23} M_{\text{sym},0}U^*_{23}$, each of which can be diagonalized by a real orthogonal matrix, respectively. The two mass matrices must satisfy the relation
\begin{eqnarray}
U^\dag_{23} M_\text{sym}U^*_{23} = I_\alpha \big(1+\frac{\epsilon}{2}\big) U^\dag_{23} M_{\text{sym},0}U^*_{23} + \frac{\epsilon}{2}
\left(
\begin{array}{ccc}
 -a & 0 & 0 \\
 0 & d-\text{Re}c & \text{Im}c \\
 0 & \text{Im}c & d+\text{Re}c
\end{array}
\right)\,.
\label{M_S&M_S0}
\end{eqnarray}
The first term of the RHS in Eq.~\eqref{M_S&M_S0} corresponds to an overall factor $I_\alpha (1+\epsilon/2)$ multiplying to the absolute neutrino masses and have no influence on the mixing parameters, and the second term may modify both masses and flavor mixing. Using the relation in Eq.~\eqref{M_S&M_S0} and taking $\epsilon$ as a small parameter, we can perturbatively obtain relations of the absolute neutrino masses between two energy scales
\begin{eqnarray}
&&m^{}_1=m^{}_{1,0}I_\alpha\left[1+\epsilon(1-c^2_{13}c^2_{12})\right] \,, \nonumber\\
&&m^{}_2=m^{}_{2,0}I_\alpha\left[1+\epsilon(1-c^2_{13}s^2_{12})\right] \,, \nonumber\\
&&m^{}_3=m^{}_{3,0}I_\alpha\left[1+\epsilon c^2_{13}\right] \,. 
\label{correction2masses}
\end{eqnarray}
We also
connect the mixing angles $\theta_{13}$, $\theta_{12}$ with $\theta_{13,0}$, $\theta_{12,0}$ as
\begin{eqnarray}
&&\theta_{13}=\theta_{13,0}-\frac{\epsilon}{2}c^{}_{13}s^{}_{13} \left[ c^2_{12}\zeta^{-\eta_{\rho}}_{31}+s^2_{12}\zeta^{-\eta_\sigma}_{32} \right] \,, \nonumber\\
&&\theta_{12}=\theta_{12,0}-\frac{\epsilon}{2}c^{}_{12}s^{}_{12}\left[s^{2}_{13}(\zeta^{-\eta_\rho}_{31}-\zeta^{-\eta_\sigma}_{32}) + c^2_{13}\zeta^{-\eta_\rho \eta_\sigma}_{21} \right]\,, 
\label{correction2th13&th12}
\end{eqnarray}
where 
\begin{eqnarray}
\zeta_{ij}=\frac{m_i-m_j}{m_i+m_j} \,,
\end{eqnarray}
and $\eta_\rho$ and $\eta_\sigma$ take only two discrete values $\pm1$. 
Note that all the mixing angles and mass parameters in the RHS of Eqs.~\eqref{correction2masses} and \eqref{correction2th13&th12} should stand for the parameters at $\Lambda_\text{FS}$ and take the subscript ``$_0$''. For those multiplied by $\epsilon$, since we list only the leading corrections, we can safely replace them with the parameters at $\Lambda_\text{EW}$ and abandon the subscript ``$_0$''.

\subsection*{$\mu$-$\tau$ anti-symmetric corrections}
\indent 

The RG-induced $\mu$-$\tau$ anti-symmetric corrections is characterized by $M_\text{asym}$. We do the following transformation for $M_\text{asym}$ and get an imaginary mass matrix
\begin{eqnarray}
U^\dag_{23} M_\text{asym}U^*_{23} = i\frac{\epsilon}{2}
\left(
\begin{array}{ccc}
 0 & -\sqrt{2}\text{Re}b & \sqrt{2}\text{Im}b \\
 -\sqrt{2}\text{Re}b & -2\text{Im}c & -2\text{Re}c \\
 \sqrt{2}\text{Im}b & -\text{Re}c & 2\text{Im}c
\end{array}
\right)\,.
\end{eqnarray}
We perturbatively diagonalize $M_\nu(\Lambda_\text{EW})=M_\text{sym}+M_\text{asym}$ around the $\mu$-$\tau$ symmetric part $M_\text{sym}$, and obtain
\begin{eqnarray}
&&\theta_{23}=45^\circ + \eta_\delta \frac{\epsilon}{2} \left(s^2_{12}\zeta^{\eta_\rho}_{31}+ c^2_{12}\zeta^{\eta_\sigma}_{32}\right)  \,,\nonumber\\
&&\delta=\eta_\delta 90^\circ+ \frac{\epsilon}{2}\left[\frac{c_{12}s_{12}}{s_{13}}\left(\zeta^{\eta_\rho}_{31}-\zeta^{\eta_\sigma}_{32}\right)+\frac{s_{13}}{c_{12}s_{12}}\left(c^4_{12}\zeta^{\eta_\sigma}_{32}-s^4_{12}\zeta^{\eta_\rho}_{31}+\zeta^{\eta_\rho\eta_\sigma}_{21}\right)\right] \,.
\label{correction2th23&delta}
\end{eqnarray}
The octant of $\theta_{23}$ depends on the sign of $\eta_\delta$, $\epsilon$, and the neutrino mass ordering (i.e., the signs of $\zeta_{31}$ and $\zeta_{32}$).
The correction $\theta_{23}-45^\circ$ should be $\lesssim10\%$ due to current neutrino oscillation data and $\gtrsim1\%$ such that it can be measured in the future experiment. The Majorana phases are also corrected by $\epsilon$
\begin{eqnarray}
&&\rho=\arg\sqrt{\eta_\rho} + \frac{\epsilon}{2} \left[ \frac{c^2_{13} c_{12} s_{12}}{s_{13}} \left(\zeta^{\eta_\sigma}_{32} - \zeta^{\eta_\rho}_{31}\right) - \frac{s_{13} s_{12}}{c_{12}} \left(\zeta^{\eta_\rho \eta_\sigma}_{21}-\zeta^{\eta_\rho}_{31} \right)\right]  \,,\nonumber\\
&&\sigma=\arg\sqrt{\eta_\sigma} + \frac{\epsilon}{2} \left[ \frac{c^2_{13} c_{12}s_{12}}{s_{13}} \left(\zeta^{\eta_\sigma}_{32} - \zeta^{\eta_\rho}_{31}\right) -\frac{s_{13} c_{12}}{s_{12}} \left(\zeta^{\eta_\rho \eta_\sigma}_{21} + \zeta^{\eta_\sigma}_{32} \right) \right] \,.
\label{correction2Majorana}
\end{eqnarray}
We also calculate the corrections to $\theta_{12,0}$, $\theta_{13,0}$ and absolute neutrino masses from the $\mu$-$\tau$ asymmetric part, and find that they are in the order $\epsilon^2$, which can be safely neglected.

Note that the RG-induced corrections to masses in Eq.~\eqref{correction2masses} and mixing parameters in Eqs.~\eqref{correction2th13&th12}, \eqref{correction2th23&delta}, \eqref{correction2Majorana} hold only for $\epsilon\zeta^{-1}_{ij}\lesssim1$. In other word, they become invalid if neutrinos have degenerate masses with $\zeta^{}_{ij}\lesssim\epsilon$.
To be compatible with experimental data, the correction $\theta_{23}-45^\circ$ should be $\lesssim10\%$, and thus the conditions $\zeta_{31}>\epsilon$ and $\zeta_{32}>\epsilon$ hold. However, we do not have such a constraint on $\zeta^{}_{21}$.
In most cases, it is very tiny due to the degenerate masses $m_1$ and $m_2$, especially in the inverted mass ordering, such that $\zeta^{}_{21}\lesssim\epsilon$ is possible. 
Later we will see that it happens in some cases of the MSSM with large $\tan\beta$. Therefore, we should turn into the perturbative calculation with degenerate eigenvalues, which can be divided into two pieces. 
We list their leading results in the following:
\begin{itemize}
\item[(A)]{$\eta_{\rho}=\eta_{\sigma}=\pm1$.}

Formulae of $m_3$, $\theta_{23}$, $\theta_{13}$, $\delta$, $\rho$ and $\sigma$  keep unchanged, but those of $m_1$, $m_2$ and $\theta_{12}$ are modified:
\begin{eqnarray}
&&m^{}_1=\frac{1}{2}(m^{}_{1,0}+m^{}_{2,0})I_\alpha\left[1+\frac{\epsilon}{2}(2-c^2_{13})- \frac{h}{2}\right] \,, \nonumber\\
&&m^{}_2=\frac{1}{2}(m^{}_{1,0}+m^{}_{2,0})I_\alpha\left[1+\frac{\epsilon}{2}(2-c^2_{13})+ \frac{h}{2}\right] \,, \nonumber\\
&&\theta_{12}=\theta_{12,0}-\frac{1}{2}\arcsin \left(\frac{\epsilon}{h} c^2_{13} \sin2\theta_{12,0} \right)\,,
\label{correction2th12_dg}
\end{eqnarray}
where
\begin{eqnarray}
h=\sqrt{(\zeta_{21}+\epsilon c^2_{13}\cos2\theta_{12,0})^2+(\epsilon c^2_{13}\sin2\theta_{12,0})^2}
\end{eqnarray}
is in the same order of $\epsilon$. 
We see that the correction to $\theta_{12,0}$ is not suppressed by $\epsilon$.

\item[(B)]{$\eta_{\rho}=-\eta_{\sigma}=\pm1$.}

This case is more complicated than (A). Only formulae of $m_3$, $\theta_{23}$ and $\theta_{13}$ in the above are valid. Those of the mass eigenvalues $m_1$ and $m_2$ are given by  
\begin{eqnarray}
&&m^{}_1=\frac{1}{2}(m^{}_{1,0}+m^{}_{2,0})I_\alpha\left[1+\frac{\epsilon}{2}(2-c^2_{13})-\sqrt{\big(\zeta_{21}+ \frac{\epsilon}{2}c^2_{13}\cos2\theta_{12,0}\big)^2+\epsilon^2s^2_{13}}\,\right] \,, \nonumber\\
&&m^{}_2=\frac{1}{2}(m^{}_{1,0}+m^{}_{2,0})I_\alpha\left[1+\frac{\epsilon}{2}(2-c^2_{13})+\sqrt{\big(\zeta_{21}+ \frac{\epsilon}{2}c^2_{13}\cos2\theta_{12,0}\big)^2+\epsilon^2s^2_{13}}\,\right] \,.
\label{degenerate_mass}
\end{eqnarray}
The leading order corrections to the other mixing parameters are expressed as
\begin{eqnarray}
&&\sin\theta_{12}=\sqrt{s^2_{12,0} c^2_\vartheta+ c^2_{12,0} s^2_\vartheta }\,,\nonumber\\
&&\tan\delta=\eta_\delta \sin2\theta_{12,0} \cot2\vartheta\,,\nonumber\\
&&\tan\rho=(\tan\theta_{12,0} \tan\vartheta)^{\eta_\rho}\,,\nonumber\\
&&\tan\sigma=(\tan\theta_{12,0} \cot\vartheta)^{\eta_\rho}\,,
\label{correction2mixing}
\end{eqnarray}
where
\begin{eqnarray}
&&\tan 2\vartheta=\frac{4\epsilon s_{13}}{2\zeta_{21}+\epsilon c^2_{13}\cos2\theta_{12,0}}\,.
\end{eqnarray}
 
\end{itemize}

\subsection{Basic features of the RG-induced $\mu$-$\tau$ reflection symmetry breaking}

The signs of the parameters $\eta_\delta$, $ \eta_\rho$ and $\eta_\sigma$ cannot be determined by the $\mu$-$\tau$ reflection symmetry. In our following discussion, we will choose $\eta_\delta=-1$, since current neutrino data hint $\delta\sim-90^\circ$ \cite{Forero:2014bxa,Gonzalez-Garcia:2014bfa}. The RG behavior in the case $\eta_\delta=1$ can be easily figured out with the help of the analytical expressions of mixing parameters and neutrino masses. Then, there are 4 different cases:
\begin{eqnarray*}
\begin{array}{lL{1cm}ll}
 \text{case\; I}  & ,  & \eta_\rho=\;\;\eta_\sigma\;=\;\;1 &;  \\
 \text{case\, II}  & ,  & \eta_\rho=\;\;\eta_\sigma\;=-1 &; \\
 \text{case III}  &  , & \eta_\rho=-\eta_\sigma=\;\;1 &; \\
 \text{case IV}  &  , & \eta_\rho=-\eta_\sigma=-1 &.
\end{array}
\end{eqnarray*}
The RG behaviors are different in these cases. From Eq.~\eqref{correction2th23&delta}, we see that $\theta_{23}$ has the largest deviation from $45^\circ$ in case II, and $\delta$ may get larger deviation from $-90^\circ$ in cases III and IV due to the enhancement of $\zeta^{-1}_{21}$. Based on the analytical calculation in the above section, we will discuss the basic features of the $\mu$-$\tau$ reflection symmetry breaking in these cases in this subsection and the numerical result in the next subsection.

The corrections to the mixing parameters are mainly dependent upon two sets of parameters: $\epsilon$ and $\zeta_{ij}$. In order to prove the $\mu$-$\tau$ symmetry breaking from the RG evolution in the future neutrino oscillation experiments, the relative corrections $\theta_{23}$ and $\delta$ in Eq.~\eqref{correction2th23&delta} should be in the order $\mathcal{O}(1\%)$ or even $\mathcal{O}(10\%)$. In the standard model, the Yukawa coupling $y_\tau$ is sufficiently small, $y_\tau\sim0.01$. If we set the flavor symmetry breaking scale $\Lambda_\text{FS}$ below but very close to the canonical seesaw scale $\Lambda_\text{FS}\sim 10^{14}$ GeV, we will get a very tiny $\epsilon\sim 10^{-5}$. Naively, we have two ways to enhance the corrections:
\begin{itemize}
\item One way is to enhance the mass parameters $\zeta^{-1}_{ij}$. We show the magnitude of $\zeta^{-1}_{ij}$ as a function of the lightest neutrino mass for both the normal mass ordering (NMO) and inverted mass ordering (IMO) in Fig.~\ref{fig:zeta}. For the lightest neutrino mass around $1$ eV,~ $\zeta^{-1}_{21}$ gains a $5\times10^4$ enhancement and $|\zeta_{31}^{-1}|$, $|\zeta_{32}^{-1}|$ gain $1.6\times10^3$ enhancements, which are large enough for significant large corrections to the $\mu$-$\tau$ reflection symmetry in the SM. However, such large masses are not compatible with the cosmological constraint. Planck sets the limit of the sum of neutrino masses less than 0.23 eV at 95\% \cite{Ade:2013zuv}, corresponding to the lightest neutrino mass $\lesssim0.07$ eV. In this case, $|\zeta^{-1}_{31}|,~|\zeta^{-1}_{32}|\lesssim10$, which are not big enough to contribute an observable correction to $\theta_{23}$ and $\delta$.
Moreover, which parameters can get large corrections are strongly dependent upon the signs of $\eta_\rho$ and $\eta_\sigma$, since the corrections are always proportional to $\zeta^{\pm \eta_\rho\eta_\sigma}_{21}$, $\zeta^{\pm \eta_\rho}_{21}$ or $\zeta^{\pm \eta_\sigma}_{21}$.

\item The other way is to extend the standard model to some new physics, such as the supersymmetric model and the more general two-Higgs doublet model \cite{Antusch:2001vn}. Since charged leptons may couple to a Higgs field different from the SM Higgs field, the magnitude of $y_\tau$ could be much larger than that in the SM. For example, in the MSSM, we have $y_\tau\sim0.01\times\tan\beta$. Given $\tan\beta=30$, $\epsilon$ can be enhanced by a factor of $30^2$, i.e., $\epsilon\sim 0.01$, and thus can reach the capability of the future neutrino oscillation experiments.  From the theoretical point of view, a large part of models have been constructed in the framework of supersymmetry since it is helpful to solve the vacuum alignment problem of flavon fields \cite{Altarelli:2005yx}. And models of generalized CP are usually realized in the supersymmetry, for instance, see \cite{Ding:2013bpa,Ding:2013hpa,Feruglio:2013hia}.
In the following, we will discuss radiative corrections in the MSSM in detail.

\begin{figure}[htbp]
\begin{minipage}[t]{0.5\linewidth}
\begin{center}
\includegraphics[width=.95\textwidth]{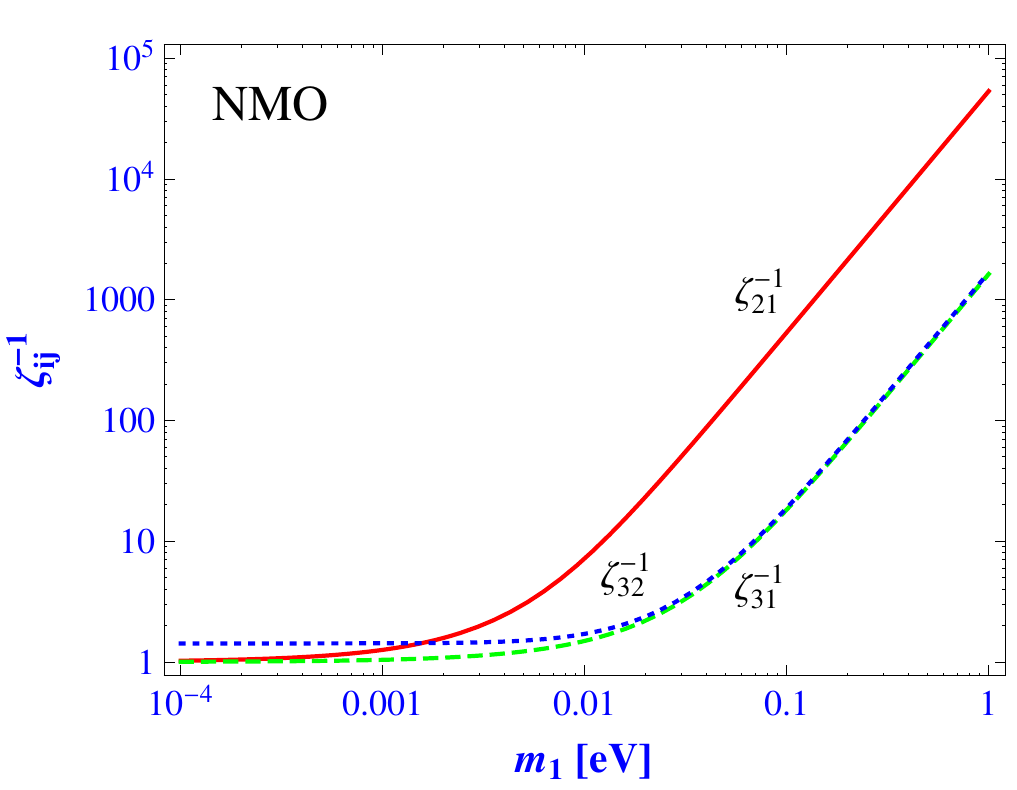}
\end{center}
\end{minipage}
\begin{minipage}[t]{0.5\linewidth}
\begin{center}
\includegraphics[width=.95\textwidth]{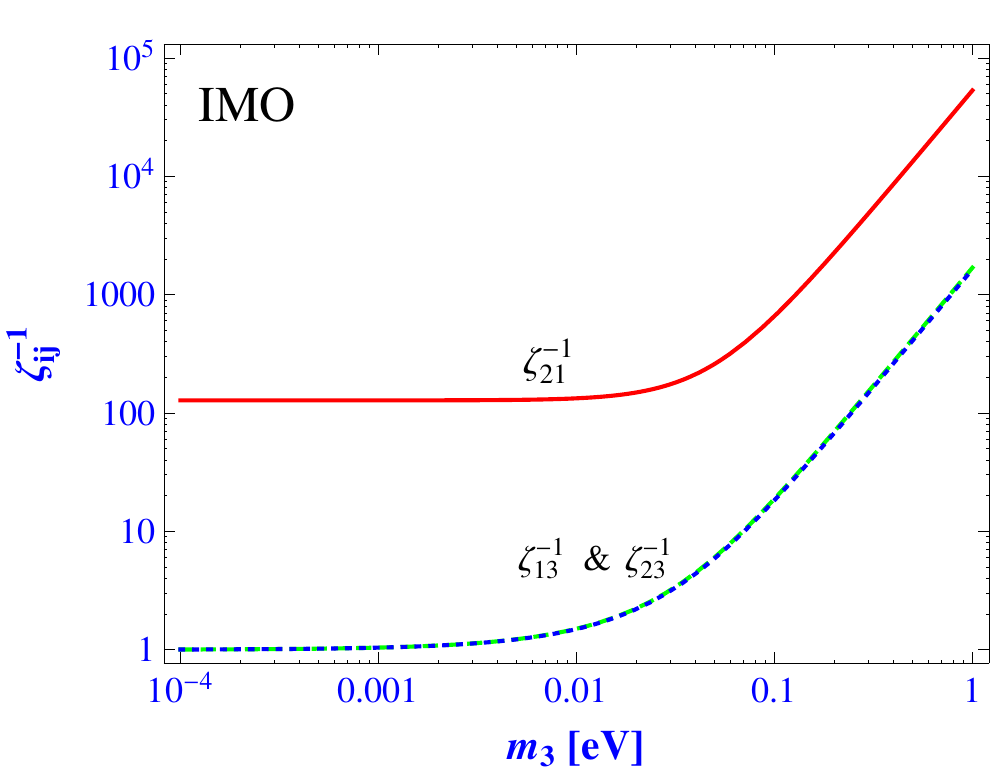}
\label{default}
\end{center}
\end{minipage}
\vspace{-.5cm}
\caption{$\zeta^{-1}_{ij}$ as a function of the lightest neutrino mass $m_1$ in the NMO or $m_3$ in the IMO. $\Delta m^2_{21}=7.50\times10^{-5}~\text{eV}^2$ and $\Delta m^2_{31}=2.457\times10^{-3}~\text{eV}^2$ for NMO ($\Delta m^2_{23}=2.449\times10^{-3}~\text{eV}^2$ for IMO) from global-fit data \cite{Gonzalez-Garcia:2014bfa} have been used as inputs. 
\label{fig:zeta}}
\end{figure}

\end{itemize}

\subsection{Numerical results}

\begin{table}[h!]
\caption{Radiative corrections of the $\mu$-$\tau$ reflection symmetry in the MSSM with $\tan\beta=10$. We fix $\theta_{23}=45^\circ$, $\delta=-90^\circ$, $\rho,\sigma=0,90^\circ$, and relax $m_1$, $\Delta m^2_{21}$, $\Delta m^2_{31}$, $\theta_{12}$, $\theta_{13}$ at the flavor symmetry breaking scale $\Lambda_\text{FS}$. After the energy scale runs down to the electroweak scale $\Lambda_\text{EW}$, we require all oscillation parameters $\Delta m^2_{21}$, $\Delta m^2_{31}$, $\theta_{23}$, $\theta_{13}$, $\theta_{12}$ should be compatible with the global-fit data in Ref.~\cite{Forero:2014bxa,Gonzalez-Garcia:2014bfa} in $3\sigma$ range.}
\label{tab1}
\begin{center}
\begin{tabular}{L{3.5cm}R{1.2cm}R{1.2cm}R{1.2cm}R{1.2cm}R{1.2cm}R{1.2cm}R{1.2cm}R{1.2cm}}
\hline\hline

 MSSM, $\tan\beta=10$ & \multicolumn{2}{c}{\quad Case I}  & \multicolumn{2}{c}{\quad Case II} & \multicolumn{2}{c}{\quad Case III}   & \multicolumn{2}{c}{\quad Case VI}    \\
NMO, $m_1\sim 0.05$ eV  & $\Lambda _{\text{FS}}$ & $\Lambda _{\text{EW}}$ & $\Lambda _{\text{FS}}$ & $\Lambda _{\text{EW}}$ & $\Lambda _{\text{FS}}$ & $\Lambda _{\text{EW}}$ & $\Lambda _{\text{FS}}$ & $\Lambda _{\text{EW}}$   \\\hline

 $m_1 [10^{-2}\text{eV}]$ & 9.62 & 4.999 & 9.62 & 4.999 & 9.62 & 4.999 & 9.62 & 4.999 \\
 $\Delta  m_{21}^2\left[10^{-5} \text{eV}^2\right]$ & 28.6 & 7.476 & 28.6 & 7.475 & 28.6 & 7.449 & 28.6 & 7.452 \\
 $\Delta  m_{31}^2\left[10^{-3} \text{eV}^2\right]$ & 8.94 & 2.406 & 8.94 & 2.406 & 8.94 & 2.406 & 8.94 & 2.406 \\
 $\theta _{12}[{}^{\circ}]$ & 33.5 & 35.96 & 33.5 & 35.97 & 33.5 & 33.51 & 33.5 & 33.50 \\
 $\theta _{13}[{}^{\circ}]$ & 8.8 & 8.837 & 8.8 & 8.801 & 8.8 & 8.826 & 8.8 & 8.813 \\
 $\theta _{23}[{}^{\circ}]$ & 45 & 45.01 & 45 & 45.25 & 45 & 45.18 & 45 & 45.08 \\
 $\delta [{}^{\circ}]$ & $-$90 & $-$90.00 & $-$90 & $-$89.99 & $-$90 & $-$91.12 & $-$90 & $-$92.51 \\
 $\rho [{}^{\circ}]$ & 0 & 0.00 & 90 & 90.06 & 0 & 0.161 & 90 & 88.79 \\
 $\sigma [{}^{\circ}]$ & 0 & 0.00 & 90 & 89.98 & 90 & 89.39 & 0 & 178.05 \\\hline\hline
 
   MSSM, $\tan\beta=10$ & \multicolumn{2}{c}{\quad Case I}  & \multicolumn{2}{c}{\quad Case II} & \multicolumn{2}{c}{\quad Case III}   & \multicolumn{2}{c}{\quad Case VI}    \\
IMO, $m_3\sim 0.05$ eV  & $\Lambda _{\text{FS}}$ & $\Lambda _{\text{EW}}$ & $\Lambda _{\text{FS}}$ & $\Lambda _{\text{EW}}$ & $\Lambda _{\text{FS}}$ & $\Lambda _{\text{EW}}$ & $\Lambda _{\text{FS}}$ & $\Lambda _{\text{EW}}$   \\\hline

 $m_3 [10^{-2}\text{eV}]$ & 9.62 & 4.995 & 9.62 & 4.995 & 9.62 & 4.995 & 9.62 & 4.995 \\
 $\Delta  m_{21}^2\left[10^{-5} \text{eV}^2\right]$ &  30.0 & 7.468 & 30.0 & 7.470 & 30.0 & 7.591 & 30.0 & 7.586 \\
 $\Delta  m_{23}^2\left[10^{-3} \text{eV}^2\right]$ & 8.94 & 2.413 & 8.94 & 2.413 & 8.94 & 2.414 & 8.94 & 2.414 \\
 $\theta _{12}~[{}^{\circ}]$ & 33.5 & 33.15 & 33.5 & 33.13 & 33.5 & 33.51 & 33.5 & 33.52 \\
 $\theta _{13}~[{}^{\circ}]$ & 8.8 & 8.763 & 8.8 & 8.799 & 8.8 & 8.774 & 8.8 & 8.789 \\
 $\theta _{23}~[{}^{\circ}]$ & 45 & 44.99 & 45 & 44.76 & 45 & 44.83 & 45 & 44.92 \\
 $\delta~ [{}^{\circ}]$ & $-$90 & $-$90.00 & $-$90 & $-$89.95 & $-$90 & $-$94.06 & $-$90 & $-$92.70 \\
 $\rho~ [{}^{\circ}]$ & 0 & 0.00 & 90 & 89.99 & 0 & 178.28 & 90 & 89.63 \\
 $\sigma~ [{}^{\circ}]$ & 0 & 0.00 & 90 & 90.07 & 90 & 87.00 & 0 & 178.33 \\\hline\hline
 
\end{tabular}
\end{center}
\end{table}

\begin{table}[h!]
\caption{Radiative corrections of the $\mu$-$\tau$ reflection symmetry in the MSSM with $\tan\beta=30$. The same requirements are taken from Table~\ref{tab1}.}
\label{tab2}
\begin{center}
\begin{tabular}{L{3.5cm}R{1.2cm}R{1.2cm}R{1.2cm}R{1.2cm}R{1.0cm}R{1.4cm}R{1.0cm}R{1.4cm}}
\hline\hline

 MSSM, $\tan\beta=30$ & \multicolumn{2}{c}{\quad Case I}  & \multicolumn{2}{c}{\quad Case II} & \multicolumn{2}{c}{\quad Case III}   & \multicolumn{2}{c}{\quad Case VI}    \\
NMO, $m_1\sim 0.05$ eV  & $\Lambda _{\text{FS}}$ & $\Lambda _{\text{EW}}$ & $\Lambda _{\text{FS}}$ & $\Lambda _{\text{EW}}$ & $\Lambda _{\text{FS}}$ & $\Lambda _{\text{EW}}$ & $\Lambda _{\text{FS}}$ & $\Lambda _{\text{EW}}$   \\\hline

 $m_1 [10^{-2} \text{eV}]$ & 9.62 & 5.045 & 9.62 & 5.045 & 10.14 & 5.036 & 10.14 & 5.035 \\
 $\Delta  m_{21}^2\left[10^{-5} \text{eV}^2\right]$ & 28.6 & 7.513 & 28.6 & 7.397 & 40.5 & 7.311 & 40.5 & 7.582 \\
 $\Delta  m_{31}^2\left[10^{-3} \text{eV}^2\right]$ & 8.94 & 2.406 & 8.94 & 2.413 & 10.16 & 2.421 & 10.16 & 2.419 \\
 $\theta _{12}~[{}^{\circ}]$ & 33.5 & 34.56 & 33.5 & 35.68 & 33.5 & 34.21 & 33.5 & 34.07 \\
 $\theta _{13}~[{}^{\circ}]$ & 8.8 & 9.178 & 8.8 & 8.802 & 8.8 & 9.112 & 8.8 & 8.982 \\
 $\theta _{23}~[{}^{\circ}]$ & 45 & 45.06 & 45 & 47.51 & 45 & 46.78 & 45 & 45.78 \\
 $\delta~ [{}^{\circ}]$ & $-$90 & $-$90.02 & $-$90 & $-$89.93 & $-$90 & $-$102.81 & $-$90 & $-$115.54 \\
 $\rho~ [{}^{\circ}]$ & 0 & 0.00 & 90 & 90.59 & 0 & 0.970 & 90 & 77.70 \\
 $\sigma~ [{}^{\circ}]$ & 0 & 179.98 & 90 & 89.78 & 90 & 82.87 & 0 & 160.27 \\\hline\hline
 
   MSSM, $\tan\beta=30$ & \multicolumn{2}{c}{\quad Case I}  & \multicolumn{2}{c}{\quad Case II} & \multicolumn{2}{c}{\quad Case III}   & \multicolumn{2}{c}{\quad Case VI}    \\
IMO, $m_3\sim 0.05$ eV  & $\Lambda _{\text{FS}}$ & $\Lambda _{\text{EW}}$ & $\Lambda _{\text{FS}}$ & $\Lambda _{\text{EW}}$ & $\Lambda _{\text{FS}}$ & $\Lambda _{\text{EW}}$ & $\Lambda _{\text{FS}}$ & $\Lambda _{\text{EW}}$   \\\hline

 $m_3 [10^{-2} \text{eV}]$ & 10.14 & 4.991 & 10.14 & 4.988 & 10.14 & 4.989 & 10.14 & 4.990 \\
 $\Delta  m_{21}^2\left[10^{-5} \text{eV}^2\right]$ & 71.8 & 7.398 & 71.8 & 7.191 & 46.2 & 7.627 & 46.2 & 7.217 \\
 $\Delta  m_{23}^2\left[10^{-3} \text{eV}^2\right]$ & 9.82 & 2.392 & 9.82 & 2.399 & 9.82 & 2.429 & 9.82 & 2.424 \\
 $\theta _{12}~[{}^{\circ}]$ & 11.4 & 33.70 & 11.4 & 33.30 & 33.5 & 35.18 & 33.5 & 35.62 \\
 $\theta _{13}~[{}^{\circ}]$ & 8.8 & 8.426 & 8.8 & 8.782 & 8.8 & 8.588 & 8.8 & 8.742 \\
 $\theta _{23}~[{}^{\circ}]$ & 45 & 44.93 & 45 & 42.57 & 45 & 43.31 & 45 & 44.21 \\
 $\delta~ [{}^{\circ}]$ & $-$90 & $-$90.00 & $-$90 & $-$89.52 & $-$90 & $-$131.38 & $-$90 & $-$120.17 \\
 $\rho~ [{}^{\circ}]$ & 0 & 179.99 & 90 & 89.91 & 0 & 162.14 & 90 & 85.00 \\
 $\sigma~ [{}^{\circ}]$ & 0 & 0.01 & 90 & 90.72 & 90 & 59.76 & 0 & 161.55 \\\hline\hline
 
\end{tabular}
\end{center}
\end{table}

\begin{table}[h!]
\caption{Radiative corrections of the $\mu$-$\tau$ reflection symmetry in the MSSM with $\tan\beta=50$. In the last two cases, there are no solutions to obtain correct values compatible with experimental data, so we use ``$-$'' instead. The same requirements are taken from Table~\ref{tab1}.}
\label{tab3}
\begin{center}
\begin{tabular}{L{3.5cm}R{1.2cm}R{1.2cm}R{1.2cm}R{1.2cm}R{1.2cm}R{1.2cm}R{1.2cm}R{1.2cm}}
\hline\hline
   MSSM, $\tan\beta=50$ & \multicolumn{2}{c}{\quad Case I}  & \multicolumn{2}{c}{\quad Case II} & \multicolumn{2}{c}{\quad Case III}   & \multicolumn{2}{c}{\quad Case VI}    \\
NMO, $m_1\sim10^{-3}~\text{eV}$  & $\Lambda _{\text{FS}}$ & $\Lambda _{\text{EW}}$ & $\Lambda _{\text{FS}}$ & $\Lambda _{\text{EW}}$ & $\Lambda _{\text{FS}}$ & $\Lambda _{\text{EW}}$ & $\Lambda _{\text{FS}}$ & $\Lambda _{\text{EW}}$   \\\hline

 $m_1 [10^{-3} \text{eV}]$ & 2.8 & 1.002 & 2.8 & 1.002 & 2.8 & 1.002 & 2.8 & 1.003 \\
 $\Delta  m_{21}^2\left[10^{-5} \text{eV}^2\right]$ & 61.08 & 7.494 & 61.08 & 7.450 & 61.08 & 7.492 & 61.08 & 7.479 \\
 $\Delta  m_{31}^2\left[10^{-3} \text{eV}^2\right]$ & 20.2 & 2.410 & 20.2 & 2.425 & 20.2 & 2.410 & 20.2 & 2.414 \\
 $\theta _{12}~[{}^{\circ}]$ & 33.5 & 34.46 & 33.5 & 34.31 & 33.5 & 34.11 & 33.5 & 34.12 \\
 $\theta _{13}~[{}^{\circ}]$ & 8.8 & 9.094 & 8.8 & 9.138 & 8.8 & 9.080 & 8.8 & 9.034 \\
 $\theta _{23}~[{}^{\circ}]$ & 45 & 46.29 & 45 & 49.50 & 45 & 46.33 & 45 & 47.12 \\
 $\delta~ [{}^{\circ}]$ & $-$90 & $-$91.85 & $-$90 & $-$79.32 & $-$90 & $-$92.51 & $-$90 & $-$88.72 \\
 $\rho~ [{}^{\circ}]$ & 0 & 178.78 & 90 & 101.76 & 90 & 88.33 & 0 & 2.196 \\
 $\sigma~ [{}^{\circ}]$ & 0 & 178.19 & 90 & 100.06 & 0 & 177.62 & 90 & 91.22\\\hline\hline
 
  MSSM, $\tan\beta=50$ & \multicolumn{2}{c}{\quad Case I}  & \multicolumn{2}{c}{\quad Case II} & \multicolumn{2}{c}{\quad Case III}   & \multicolumn{2}{c}{\quad Case VI}    \\
IMO, $m_3\sim10^{-3}~\text{eV}$  & $\Lambda _{\text{FS}}$ & $\Lambda _{\text{EW}}$ & $\Lambda _{\text{FS}}$ & $\Lambda _{\text{EW}}$ & $\Lambda _{\text{FS}}$ & $\Lambda _{\text{EW}}$ & $\Lambda _{\text{FS}}$ & $\Lambda _{\text{EW}}$   \\\hline

 $m_3 [10^{-3} \text{eV}]$ & 2.9 & 0.999 & 2.9 & 0.998 & ----- & ----- & ----- & ----- \\
 $\Delta  m_{21}^2\left[10^{-5} \text{eV}^2\right]$ & 238.0 & 7.369 & 238.0 & 7.394 & ----- & ----- & ----- & ----- \\
 $\Delta  m_{31}^2\left[10^{-3} \text{eV}^2\right]$ & 20.2 & 2.424 & 20.2 & 2.424 & ----- & ----- & ----- & ----- \\
 $\theta _{12}~[{}^{\circ}]$ & 6.6 & 34.02 & 6.6 & 33.79 & ----- & ----- & ----- & ----- \\
 $\theta _{13}~[{}^{\circ}]$ & 8.8 & 8.534 & 8.8 & 8.555 & ----- & ----- & ----- & ----- \\
 $\theta _{23}~[{}^{\circ}]$ & 45 & 43.41 & 45 & 43.28 & ----- & ----- & ----- & ----- \\
 $\delta~ [{}^{\circ}]$ & $-$90 & $-$89.82 & $-$90 & $-$89.79 & ----- & ----- & ----- & ----- \\
 $\rho~ [{}^{\circ}]$ & 0 & 179.84 & 90 & 89.83 & ----- & ----- & ----- & ----- \\
 $\sigma~ [{}^{\circ}]$ & 0 & 0.34 & 90 & 90.38 & ----- & ----- & ----- & ----- \\\hline\hline
\end{tabular}
\end{center}
\end{table}

We perform the numerical illustration for RG corrections in the MSSM. 
We fix $\theta_{23}=45^\circ$, $\delta=-90^\circ$, $\rho,\sigma=0,90^\circ$, and keep $m_1$, $\Delta m^2_{21}$, $\Delta m^2_{31}$, $\theta_{12}$, $\theta_{13}$ as varying numbers at the flavor symmetry breaking scale $\Lambda_\text{FS}$. After the energy scale runs down to the electroweak scale $\Lambda_\text{EW}$, we require all oscillation parameters $\Delta m^2_{21}$, $\Delta m^2_{31}$, $\theta_{23}$, $\theta_{13}$, $\theta_{12}$ should be compatible with the global-fit data in Ref.~\cite{Forero:2014bxa,Gonzalez-Garcia:2014bfa} in $3\sigma$ range.
In order to see to what extent the $\mu$-$\tau$ reflection symmetry is broken, we set $\tan\beta=10,~30,~50$, with the results shown in Tables~\ref{tab1}, \ref{tab2}, \ref{tab3}, respectively. 
\begin{itemize}
\item For $\tan\beta=10$, the RG running  effect is very weak, and the corrections to $\theta_{23}$ and the CP-violating phases are less than $0.3^\circ$ and $5^\circ$, respectively. Since $\epsilon<0$ and we have chosen $\eta_\delta=-1$, the octant of $\theta_{23}$ is dependent upon the neutrino mass ordering. As shown in Eq.~\eqref{correction2th23&delta}, the NMO corresponds to $\zeta_{31},\zeta_{32}>0$, and $\theta_{23}$ belongs to the second octant ($\theta_{23}>45^\circ$). The deviation of $\delta$ from $-90^\circ$ is in general dependent upon the cancellation of $\zeta_{ij}$. In cases III and IV, since $\eta_\rho=-\eta_\sigma$ and $\zeta^{-1}_{21}\gg\zeta^{-1}_{31},\zeta^{-1}_{32}$ holds in most cases, the corrections to $\delta$ are negative, and much larger than those in cases I and II. RG behaviors of the Majorana phases are similar to those of the Dirac phase.

\item For $\tan\beta=30$, the deviation of $\theta_{23}$ can maximally reach $2.5^\circ$. We remind that $\zeta_{21}$ and $\epsilon$ are in the same order in this scenario, such that the corrections to some of the mixing parameters can be very large. In cases I and II, a large RG correction can push them to be compatible with data even if $\theta_{12}$ is sufficiently small $\simeq 11.4^\circ$ at $\Lambda_\text{FS}$, which is consistent with Eq.~\eqref{correction2th12_dg}. 
In cases III and IV, the deviation of the Dirac phase $\delta$ can be as large as $30^\circ$ to $40^\circ$, and the other CP-violating phases also acquire large corrections, which are confirmed in Eq.~\eqref{correction2mixing}. 

\item For $\tan\beta=50$, the large $\epsilon$ leads to a large correction to the neutrino mass-squared difference $\Delta m^2_{21}$, which is not consistent with neutrino oscillation data, expect that the lightest neutrino mass is small enough. In this scenario, we decrease the output lightest neutrino mass to $\simeq 10^{-3}$ eV. 
The largest deviation of $\theta_{23}$ is around $4.5^\circ$, and $\theta_{12}$ at $\Lambda_\text{FS}$ can be as small as $6.6^\circ$, smaller than $\theta_{13}$.
There is no solution in cases III and IV for the IMO due to the large correction to $\Delta m^2_{21}$. 
\end{itemize}
In short, we have found that radiative corrections in the MSSM have definite directions: $\theta_{23}>45^\circ$ in the NMO and $\theta_{23}<45^\circ$ in the IMO, and the large correction to $\delta$ always results in $\delta<-90^\circ$. It is interesting to compare these results with current global analysis of neutrino oscillation data, where the best-fit values for $\theta_{23}$ and $\delta$ are
\begin{eqnarray}
&&\theta_{23}=
\left\{\begin{array}{r}
48.9^\circ\\
49.2^\circ
\end{array} \right.,\quad 
\delta=
\left\{\begin{array}{r}
-119^\circ  \\
-94^\circ 
\end{array} \right.
\begin{array}{l}
 \text{for NMO} \\
 \text{for IMO} 
\end{array} \quad \text{from \cite{Forero:2014bxa}}\,,  \nonumber\\
&&\theta_{23}=
\left\{\begin{array}{c}
42.3^\circ\\
49.5^\circ
\end{array} \right.,\quad 
\delta=
\left\{\begin{array}{r}
-54^\circ  \\
-106^\circ 
\end{array} \right.
\begin{array}{l}
 \text{for NMO} \\
 \text{for IMO} 
\end{array} \quad \text{from \cite{Gonzalez-Garcia:2014bfa}}\,. \nonumber
\end{eqnarray}
The future neutrino oscillation experiments will determine the neutrino mass order, the octant of $\theta_{23}$ and the Dirac phase $\delta$. After that we will have more concrete ideas for the $\mu$-$\tau$ reflection symmetry and its breaking. For example, if the NMO, $\theta_{23}<45^\circ$ and $\delta$ close to but $<-90^\circ$ are finally verified, we would conclude that the $\mu$-$\tau$ reflection symmetry is an approximate symmetry, but there should be another mechanism beyond radiative corrections in the MSSM to break it.

We have also performed the numerical illustration in the SM. Since $\epsilon>0$ in the SM, we arrive at $\theta_{23}<45^\circ$ in the NMO and $\theta_{23}>45^\circ$ in the IMO at $\Lambda_\text{EW}$.
However, to obtain sufficiently large correction to $\theta_{23}$ around $0.5^\circ$ to $5^\circ$, the lightest neutrino mass should be around 1 eV. For $m_i\lesssim0.07$ eV, the RG-induced $\mu$-$\tau$ reflection symmetry breaking is unobservable in the SM.

\section{Conclusion}

The $\mu$-$\tau$ reflection symmetry can be regarded as an approximate symmetry due to its consistence with current neutrino experimental data and convenience for model building. In this paper, we have considered some basic properties of the $\mu$-$\tau$ reflection symmetry and its RG-induced correction. We assume the $\mu$-$\tau$ reflection symmetry as a remnant symmetry from an underlying family symmetry broken at sufficiently high energy scale. After the energy scale runs down to the electroweak scale, the $\mu$-$\tau$ reflection symmetry must be broken due to the radiation corrections.  

We prove that the exact $\mu$-$\tau$ reflection symmetry guarantees $\theta_{23}=45^\circ$, $\delta=\pm90^\circ$ and $\rho,\sigma=0,90^\circ$. 
In the seesaw mechanism, the $\mu$-$\tau$ reflection transformation can be extended to the sector of right-handed neutrinos. After right-handed neutrinos are integrated out, the left-handed neutrinos acquire masses, and the $\mu$-$\tau$ reflection symmetry is still preserved.

The radiative corrections to the $\mu$-$\tau$ reflection symmetry can be divided into two parts: $\mu$-$\tau$ symmetric and anti-symmetric parts.
The $\mu$-$\tau$ symmetric part modifies the values of neutrino masses $m_i$, the corresponding mass-squared differences $\Delta m^2_{21}$ and $\Delta m^2_{31}$, and the mixing angles $\theta_{12}$, $\theta_{13}$, but preserves the mixing angle $\theta_{23}=45^\circ$, the Dirac phase $\delta=\pm90^\circ$ and Majorana phases $\rho, \sigma =0, 90^\circ$ explicitly. 
The $\mu$-$\tau$ anti-symmetric part violates the $\mu$-$\tau$ reflection symmetry. As a consequence, $\theta_{23}$ and $\delta$ deviate from $45^\circ$ and $\pm90^\circ$, respectively, both in the order of $\epsilon$. The Majorana phases $\rho,\sigma$ also gain corrections at the same level, but the corrections to absolute neutrino masses $m_i$ and mixing angles $\theta_{12},\theta_{13}$ are in general very tiny, in the order $\epsilon^2$.

We point out that the RG-induced $\mu$-$\tau$ reflection symmetry breaking is negligibly small in the SM, but may be sizable in the MSSM depending on $\tan\beta$. The octant of $\theta_{23}$ after radiative corrections is determined by the neutrino mass ordering. $\theta_{23}>45^\circ$ for the NMO and $\theta_{23}<45^\circ$ for the IMO in the MSSM if $\delta$ takes the value around its current best-fit result $-90^\circ$. The corrections to all CP-violating phases $\delta,\rho,\sigma$ in case III and IV are negative and much greater than those in cases I and II for both NMO and IMO. For large $\tan\beta$, current data of $\theta_{12}$ could be an accidental result from a small angle at the flavor symmetry breaking scale, even smaller than $\theta_{13}$. Since the deviations of $\theta_{23}$ and $\delta$ have definite directions, they can be tested in the future neutrino oscillation experiments.

\vspace{0.5cm}

\section*{Acknowledgements}
The author would like to thank Prof. Zhi-zhong Xing for reading this manuscript and helpful suggestions. This work is supported in part by the National Natural Science Foundation of China under Grant No. 11135009.


\newpage
\begin{thebibliography}{99}

\bibitem{PDG}
  K.~A.~Olive {\it et al.}  [Particle Data Group Collaboration],
  Chin.\ Phys.\ C {\bf 38}, 090001 (2014).
  
\bibitem{mutau0}
  T.~Fukuyama and H.~Nishiura,
  hep-ph/9702253;
  R.~N.~Mohapatra and S.~Nussinov,
  Phys.\ Rev.\ D {\bf 60}, 013002 (1999)
  [hep-ph/9809415];
  E.~Ma and M.~Raidal,
  Phys.\ Rev.\ Lett.\  {\bf 87}, 011802 (2001)
  [Erratum-ibid.\  {\bf 87}, 159901 (2001)]
  [hep-ph/0102255];
  C.~S.~Lam,
  Phys.\ Lett.\ B {\bf 507}, 214 (2001)
  [hep-ph/0104116];
  K.~R.~S.~Balaji, W.~Grimus and T.~Schwetz,
  Phys.\ Lett.\ B {\bf 508}, 301 (2001)
  [hep-ph/0104035];
  E.~Ma,
  Phys.\ Rev.\ D {\bf 66}, 117301 (2002)
  [hep-ph/0207352].

\bibitem{BM}
  F.~Vissani,
  hep-ph/9708483;
  V.~D.~Barger, S.~Pakvasa, T.~J.~Weiler and K.~Whisnant,
  Phys.\ Lett.\ B {\bf 437}, 107 (1998)
  [hep-ph/9806387].

\bibitem{TBM}  
  P.~F.~Harrison, D.~H.~Perkins and W.~G.~Scott,
  Phys.\ Lett.\ B {\bf 530}, 167 (2002)
  [hep-ph/0202074];
  Z.~Z.~Xing,
  Phys.\ Lett.\ B {\bf 533}, 85 (2002)
  [hep-ph/0204049];
  P.~F.~Harrison and W.~G.~Scott,
  Phys.\ Lett.\ B {\bf 535}, 163 (2002)
  [hep-ph/0203209];
  X.~G.~He and A.~Zee,
  Phys.\ Lett.\ B {\bf 560}, 87 (2003)
  [hep-ph/0301092].
  
\bibitem{mutau1}
  K.~S.~Babu, E.~Ma and J.~W.~F.~Valle,
  Phys.\ Lett.\ B {\bf 552}, 207 (2003)
  [hep-ph/0206292];
  E.~Ma,
  Mod.\ Phys.\ Lett.\ A {\bf 17}, 2361 (2002)
  [hep-ph/0211393].
  
\bibitem{Grimus:2003yn} 
  W.~Grimus and L.~Lavoura,
  Phys.\ Lett.\ B {\bf 579}, 113 (2004)
  [hep-ph/0305309].
  
\bibitem{Feruglio:2012cw}
  F.~Feruglio, C.~Hagedorn and R.~Ziegler,
  JHEP {\bf 1307}, 027 (2013)
  [arXiv:1211.5560 [hep-ph]].
  
\bibitem{Holthausen:2012dk}
  M.~Holthausen, M.~Lindner and M.~A.~Schmidt,
  JHEP {\bf 1304}, 122 (2013)  [arXiv:1211.6953 [hep-ph]].
  
\bibitem{Harrison:2002et} 
  P.~F.~Harrison and W.~G.~Scott,
  Phys.\ Lett.\ B {\bf 547}, 219 (2002)
  [hep-ph/0210197].
  
\bibitem{Farzan:2006vj} 
  Y.~Farzan and A.~Y.~Smirnov,
  JHEP {\bf 0701}, 059 (2007)
  [hep-ph/0610337].

\bibitem{Zhou:2012zj} 
  S.~Zhou,
  arXiv:1205.0761 [hep-ph];
  S.~Gupta, A.~S.~Joshipura and K.~M.~Patel,
  JHEP {\bf 1309}, 035 (2013)
  [arXiv:1301.7130 [hep-ph]].
  
  
\bibitem{DYB}
  F.~P.~An {\it et al.}  [Daya Bay Collaboration],
  Phys.\ Rev.\ Lett.\  {\bf 108}, 171803 (2012)
  [arXiv:1203.1669 [hep-ex]].
  
\bibitem{T2K}
  K.~Abe {\it et al.}  [T2K Collaboration],
  Phys.\ Rev.\ Lett.\  {\bf 112}, 061802 (2014)
  [arXiv:1311.4750 [hep-ex]].
  
\bibitem{Xing:2014zka} 
  Z.~Z.~Xing and S.~Zhou,
  Phys.\ Lett.\ B {\bf 737}, 196 (2014)
  [arXiv:1404.7021 [hep-ph]];
    
\bibitem{TM}
  Z.~Z.~Xing and S.~Zhou, 
  Phys.\ Lett.\ B {\bf 653}, 278 (2007)
  [hep-ph/0607302];
  C.~S.~Lam,
  Phys.\ Rev.\ D {\bf 74}, 113004 (2006)
  [hep-ph/0611017];
  C.~H.~Albright and W.~Rodejohann,
  Eur.\ Phys.\ J.\ C {\bf 62}, 599 (2009)
  [arXiv:0812.0436 [hep-ph]].
  C.~H.~Albright, A.~Dueck and W.~Rodejohann,
  Eur.\ Phys.\ J.\ C {\bf 70}, 1099 (2010)
  [arXiv:1004.2798 [hep-ph]];
  Z.~h.~Zhao,
  arXiv:1405.3022 [hep-ph].

\bibitem{more}
  I.~Aizawa, T.~Kitabayashi and M.~Yasue,
  Phys.\ Rev.\ D {\bf 72}, 055014 (2005)
  [hep-ph/0504172];
  I.~Aizawa, T.~Kitabayashi and M.~Yasue,
  Nucl.\ Phys.\ B {\bf 728}, 220 (2005)
  [hep-ph/0507332];
  T.~Baba and M.~Yasue,
  Phys.\ Rev.\ D {\bf 75}, 055001 (2007)
  [hep-ph/0612034];
  T.~Baba and M.~Yasue,
  Phys.\ Rev.\ D {\bf 77}, 075008 (2008)
  [arXiv:0710.2713 [hep-ph]].

\bibitem{Ding:2013bpa} 
  G.~J.~Ding, S.~F.~King and A.~J.~Stuart,
  JHEP {\bf 1312}, 006 (2013)
  [arXiv:1307.4212 [hep-ph]].
  
\bibitem{Ding:2013hpa} 
  G.~J.~Ding, S.~F.~King, C.~Luhn and A.~J.~Stuart,
  JHEP {\bf 1305}, 084 (2013)
  [arXiv:1303.6180 [hep-ph]].
  
\bibitem{D48}
  G.~J.~Ding and Y.~L.~Zhou,
  arXiv:1312.5222 [hep-ph];
  G.~J.~Ding and Y.~L.~Zhou,
  JHEP {\bf 1406}, 023 (2014)
  [arXiv:1404.0592 [hep-ph]].

\bibitem{D96}  
  G.~J.~Ding and S.~F.~King,
  Phys.\ Rev.\ D {\bf 89}, 093020 (2014)
  [arXiv:1403.5846 [hep-ph]].

\bibitem{Feruglio:2013hia}
  F.~Feruglio, C.~Hagedorn and R.~Ziegler,
  Eur.\ Phys.\ J.\ C {\bf 74}, 2753 (2014)
  [arXiv:1303.7178 [hep-ph]].
  
\bibitem{models}
  W.~Grimus and L.~Lavoura,
  Fortsch.\ Phys.\  {\bf 61}, 535 (2013)
  [arXiv:1207.1678 [hep-ph]].
  R.~N.~Mohapatra and C.~C.~Nishi,
  Phys.\ Rev.\ D {\bf 86}, 073007 (2012)
  [arXiv:1208.2875 [hep-ph]].
  
\bibitem{FL}
  R.~Friedberg and T.~D.~Lee,
  HEP\&NP {\bf 30}, 591 (2006),
  hep-ph/0606071;
  Z.~Z.~Xing, H.~Zhang and S.~Zhou,
  Phys.\ Lett.\ B {\bf 641}, 189 (2006)
  [hep-ph/0607091].
  
\bibitem{RGE1}
  P.~H.~Chankowski and Z.~Pluciennik,
  Phys.\ Lett.\ B {\bf 316}, 312 (1993)
  [hep-ph/9306333];
  K.~S.~Babu, C.~N.~Leung and J.~T.~Pantaleone,
  Phys.\ Lett.\ B {\bf 319}, 191 (1993)
  [hep-ph/9309223];
  S.~Antusch, M.~Drees, J.~Kersten, M.~Lindner and M.~Ratz,
  Phys.\ Lett.\ B {\bf 519}, 238 (2001)
  [hep-ph/0108005].
  
\bibitem{RGE2}
  N.~Haba, N.~Okamura and M.~Sugiura,
  Prog.\ Theor.\ Phys.\  {\bf 103}, 367 (2000)
  [hep-ph/9810471];
  J.~A.~Casas, J.~R.~Espinosa, A.~Ibarra and I.~Navarro,
  Nucl.\ Phys.\ B {\bf 573}, 652 (2000)
  [hep-ph/9910420];
  S.~Antusch, J.~Kersten, M.~Lindner and M.~Ratz,
  Nucl.\ Phys.\ B {\bf 674}, 401 (2003)
  [hep-ph/0305273];
  S.~Antusch, J.~Kersten, M.~Lindner, M.~Ratz and M.~A.~Schmidt,
  JHEP {\bf 0503}, 024 (2005)
  [hep-ph/0501272];
  J.~w.~Mei,
  Phys.\ Rev.\ D {\bf 71}, 073012 (2005)
  [hep-ph/0502015].
  
\bibitem{RGE3}
  J.~R.~Ellis and S.~Lola,
  Phys.\ Lett.\ B {\bf 458}, 310 (1999)
  [hep-ph/9904279];
  H.~Fritzsch and Z.~z.~Xing,
  Prog.\ Part.\ Nucl.\ Phys.\  {\bf 45}, 1 (2000)
  [hep-ph/9912358];
  Z.~Z.~Xing,
  Phys.\ Rev.\ D {\bf 63}, 057301 (2001)
  [hep-ph/0011217].
  J.~w.~Mei and Z.~Z.~Xing,
  Phys.\ Rev.\ D {\bf 69}, 073003 (2004)
  [hep-ph/0312167].
  
\bibitem{Antusch:2001vn} 
  S.~Antusch, M.~Drees, J.~Kersten, M.~Lindner and M.~Ratz,
  Phys.\ Lett.\ B {\bf 525}, 130 (2002)
  [hep-ph/0110366].
  
\bibitem{Luo:2014upa} 
  S.~Luo and Z.~Z.~Xing,
  arXiv:1408.5005 [hep-ph].
  
\bibitem{Toshev:1991ku} 
  S.~Toshev,
  Mod.\ Phys.\ Lett.\ A {\bf 6}, 455 (1991).
  
\bibitem{mattereffect}
  Z.~Z.~Xing and Y.~L.~Zhou,
  Phys.\ Lett.\ B {\bf 693}, 584 (2010)
  [arXiv:1008.4906 [hep-ph]];
  Y.~L.~Zhou,
  Phys.\ Rev.\ D {\bf 84}, 113012 (2011)
  [arXiv:1110.5023 [hep-ph]].
  
\bibitem{Forero:2014bxa} 
  D.~V.~Forero, M.~Tortola and J.~W.~F.~Valle,
  arXiv:1405.7540 [hep-ph].  
  
\bibitem{Gonzalez-Garcia:2014bfa} 
  M.~C.~Gonzalez-Garcia, M.~Maltoni and T.~Schwetz,
  arXiv:1409.5439 [hep-ph].
  
\bibitem{Ade:2013zuv} 
  P.~A.~R.~Ade {\it et al.}  [Planck Collaboration],
  arXiv:1303.5076 [astro-ph.CO].

\bibitem{Altarelli:2005yx} 
  G.~Altarelli and F.~Feruglio,
  Nucl.\ Phys.\ B {\bf 741}, 215 (2006)
  [hep-ph/0512103].



\end{thebibliography}
\end{document}